\documentclass[conference]{IEEEtran}

\usepackage{subfigure}
\usepackage{booktabs}
\usepackage{amsmath,amssymb,amsfonts}
\usepackage{algorithmic}
\usepackage{graphicx}

\usepackage{textcomp}
\usepackage{xcolor}
\usepackage{makecell}

\def\BibTeX{{\rm B\kern-.05em{\sc i\kern-.025em b}\kern-.08em
    T\kern-.1667em\lower.7ex\hbox{E}\kern-.125emX}}
\begin{document}

%\title{Music Clips Correlation Network for Similarity between Music Structure and Military Strategy \\

\title{A Similarity Network for Correlating Musical Structure to Military Strategy \\

}
% \iffalse
% double blank means there is no author info in this version.

\author{
\IEEEauthorblockN{Yiwen Zhang}
andyzyw2001@gmail.com\\
\makecell[c]{Faculty of Science and Technology\\United International College}
\IEEEauthorblockA{\textit{BNU-HKBU}\\
Zhuhai, China \\
}
\and
\IEEEauthorblockN{Hui Zhang}
amyzhang@uic.edu.cn\\
\makecell[c]{Faculty of Science and Technology\\United International College}
\IEEEauthorblockA{\textit{BNU-HKBU}\\
Zhuhai, China}
\and
\IEEEauthorblockN{Fanqin Meng}
echofqmeng@bnu.edu.cn\\
\makecell[c]{Faculty of Media Arts and Design\\United International College}
\IEEEauthorblockA{\textit{BNU-HKBU}\\
Zhuhai, China}
}
% \fi

\maketitle

\begin{abstract}

Music perception, a multi-sensory process based on the synesthesia effect, is an essential component of music aesthetic education. Understanding music structure helps both perception and aesthetic education. Music structure incorporates a range of information, the coordination of which forms the melody, just as different military actions cooperate to produce a military strategy. However, there are a few ways for assessing music perception from the perspectives of system operation and information management. In this paper, we explore the similarities between music structure and military strategy while creating the Music Clips Correlation Network (MCCN) based on Mel-frequency Cepstral Coefficients (MFCCs). The inspiration comes from the comparison between a concert conductor's musical score and a military war commander's sand table exercise. Specifically, we create MCCNs for various kinds of war movie soundtracks, then relate military tactics (Sun Tzu's Art of War, etc.) and political institutions to military operations networks. Our primary findings suggest a few similarities, implying that music perception and aesthetic education can be approached from a military strategy and management perspective through this interdisciplinary research. Similarly, we can discover similarities between the art of military scheming and the art of musical structure based on network analysis in order to facilitate the understanding of the relationship between technology and art.

\end{abstract}

\begin{IEEEkeywords}
graph neural network, music perception, musical aesthetics, synesthesia, music network, music visualization
\end{IEEEkeywords}

\section{Introduction}

In the contemporary era, the task of "comprehending music" poses a challenge for individuals of all age groups. It is a common occurrence in our lives to witness instances where a young person attending a concert may exclaim, "I wish there were words to aid my understanding, but after listening for hours, I remain bewildered." In addition, when recommending classical music to our older generation, they often decline, citing their inability to grasp it. In response to such dilemmas, Prof. Chou \cite{Zhou2007} introduced the concept of "music does not necessarily require understanding". Within the realm of music psychology, the concept of "synesthesia" comes to the fore, signifying that when individuals listen to music, they instinctively connect it with other sensory experiences, such as fragrances and colors. Building upon this notion of musical synesthesia, we can harness the sensory experiences of other senses to assist individuals in perceiving music. However, it is insufficient to merely "sense" it; rather, there is a need for systematic sensory engagement. For instance, one may hear every note within Beethoven's Symphony of Fate, yet still fail to truly experience it, especially in the context of a more extended symphony. This underscores the necessity for a multi-sensory and systematic approach to music perception. Such an approach has the potential to help individuals in cultivating their associative faculties, enhancing their music perception, and facilitating both music aesthetics and music education.

In this paper, we present a systematic approach to music perception using artificial intelligence methods. Through this approach, we unveil systematic parallels between music structure and military strategy. Our exploration of these systematic similarities is inspired by the resemblance between a symphony conductor's score and a military sand table exercise. Just as multiple musical voices harmonize to create a beautiful piece of music, the synchronized actions of various military units can formulate a comprehensive military strategy. If we regard both a captivating musical composition and a successful military strategy as "achievements," it becomes evident that the underlying factor in achieving such outcomes is a coherent system comprised of interconnected components. One could even draw a parallel between the conductor of an orchestra and the commander of a military campaign, as both play a pivotal role in directing various elements towards the ultimate goal, be it a triumphant concert or a victorious battle. With this perspective in mind, let us elucidate the concepts of musical structure and military strategy. \textbf{\textit{Music structure}} encompasses composition, harmony, melody, and rhythm, offering essential insights into the evolution of music. Grasping the intricacies of musical structure entails comprehending the inner workings of musical elements and the coordination patterns akin to \textbf{\textit{military strategy}}. The latter focuses on the integration and efficient management of information and resources, which can provide valuable perspectives for enhancing music perception.

To further delve into the systematic aspects of music structure and draw comparisons between the systematic nature of military strategy and music structure, we first need to understand the applications of AI within both systems. In recent studies, the significance of advanced AI techniques in the military area, particularly for strategic decision-making and knowledge management known as OODA (Observation, Orientation, Decision, and Action) \cite{mn4}, has been highlighted. The analysis of music stimuli's impact on psychological empathy by Zhang et al. \cite{mc1} highlighted music's STR (spectral, temporal, and rhythmic) characteristics, which can be shown by MFCC. Both the systematic features (OODA and STR) for music structure and military strategy can be analyzed using a graph neural network. So we can regard the military strategy system as a case to explore some similarities between systems and learn music in a systematic way. 

In addition, we can find some useful AI methods in these two systems.  Nardelli et al. \cite{mt2} explored network representations of musical structure and offered frameworks to analyze musical objects. It is necessary to enrich the music network nodes with comprehensive features and incorporate the military nature. Previous research by Song et al. \cite{mn2} focused on modeling and simulating military systems using complex networks, resulting in the BA-NW-C2NM military model based on traditional networks. Similarly, He et al. \cite{mn1} explored information warfare systems using complex network topology models, specifically analyzing traditional operational network topology models such as Random Tree Network (RTN) and Random Apollo Network (RAN) within the information warfare network topology model. These studies identified four distinct military operational networks: RTN, RAN, System of Systems Network (SOS), and BA-NW-C2NM Network (BA). These four military system networks have distinctive characteristics in terms of construction and response to attacks. We will correlate data networks with traditional cultural political institutions and military strategies based on the algorithms used for network construction and the visual structural features. Additionally, we will employ MFCC to construct various types of music networks, thereby exploring the similarities between the two network systems.

\begin{figure}
    \centering
    \includegraphics[scale=0.5]{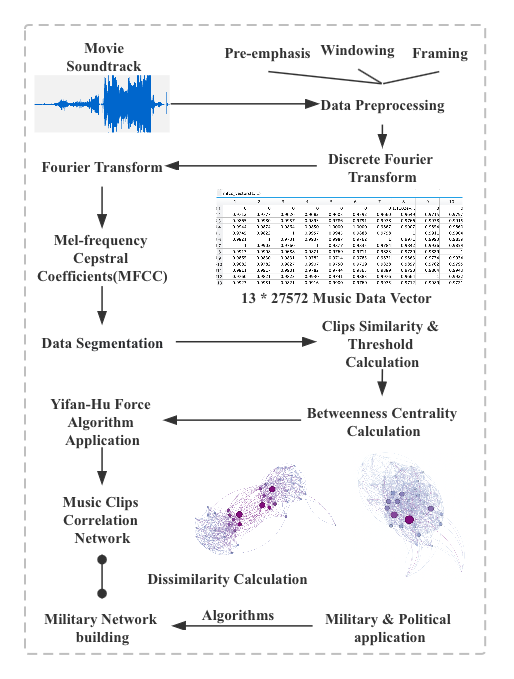}
    \label{f1}
    \caption{The techniques and steps used to study the "systematicness
" of music are presented in this paper.}
\end{figure}
\section{Proposed Work}

In this section, we will elucidate the process of constructing a music network using computational methods and imbuing it with visualized systematic attributes through various algorithms. Subsequently, we will employ the military system as a case study to explore the similarities between military strategy and music structure, thereby revealing the analogical aspects between the systems of music structure and military strategy. In the construction of the military system, we will integrate ancient political institutions and Sun Tzu's Art of War into the network framework, thereby accentuating the distinctive features of the military network.

\subsection{Music Clips Correlation Network Model Training}

\textbf{MFCC Calculation.} According to Weber's Law in Perceptual Psychology, the human just noticeable difference in signal values maintains a constant ratio, denoted as $JND / S = K$. When we employ Mel-frequency cepstral coefficients (MFCC) as the basis for music analysis, we convert frequency units from Hertz to MFCC. As a result, our computed data enhances the responsiveness of the human ear to music, better reflecting human understanding of music. We apply pre-emphasis and windowing to the music since the audio signal can be considered flat during brief time intervals. Subsequently, we utilize the Discrete Fourier Transform to analyze the energy distribution of the audio across multiple frequency bands. Finally, the frequencies obtained from the Fourier transform are mapped to the Mel scale, resulting in 13 rows of MFCC vectors for the music, where each column vector represents a single frame of music content.

\textbf{MCCN Calculation.} Following this, we construct the MCCN network as follows: \textbf{a)} We balance accuracy, robustness, and computational complexity before segmenting the MFCC vector to create MFCC clips with a consistent music length. The musical features contained in each segment are stored in a node in the form of an MFCC vector, forming a music network; \textbf{b)} We calculate the cosine similarity $similarity = \cos \theta = \frac{A\cdot B}{\left \| A \right \| \left \| B \right \| }$ between two clips of the same music to determine the similarity value; \textbf{c)} We calculate the median of the similarity values among all clips of the same music as the threshold for the music network edge; \textbf{d)} We establish a connection between two nodes if their similarity value is greater than or equal to the threshold.

\textbf{Network Algorithm Optimization.} After constructing the MCCN network, we employ the following processes to visualize and accentuate its military characteristics: \textbf{a)} We utilize the Yifan Hu algorithm, widely recognized for its effectiveness in visualizing complex networks. The Yifan Hu algorithm is a force-directed layout method employed for graph visualization. It commences by initially assigning random positions to the nodes within the graph. Subsequently, it computes repulsive forces between nodes, taking into account their distances, as well as attractive forces along edges. Node positions are then updated iteratively, with nodes gradually shifting towards positions that minimize an energy function. This iterative process continues until convergence is achieved or a pre-defined maximum number of iterations is reached. The resultant layout offers a visually informative depiction of the graph's structure, rendering it a valuable tool for network visualization across diverse applications. \textbf{b)} Based on their Betweenness Centrality, nodes are categorized as core, important, and general. The Betweenness Centrality metric assesses a node's ability to transmit information and act as an intermediary connecting other nodes. It identifies nodes that play a substantial role in military information communication and resource allocation. This ranking system allows us to gain a clearer understanding of the nodes' roles and capabilities. 

\begin{figure}
\centering
\subfigure(a){
\includegraphics[scale=0.05]{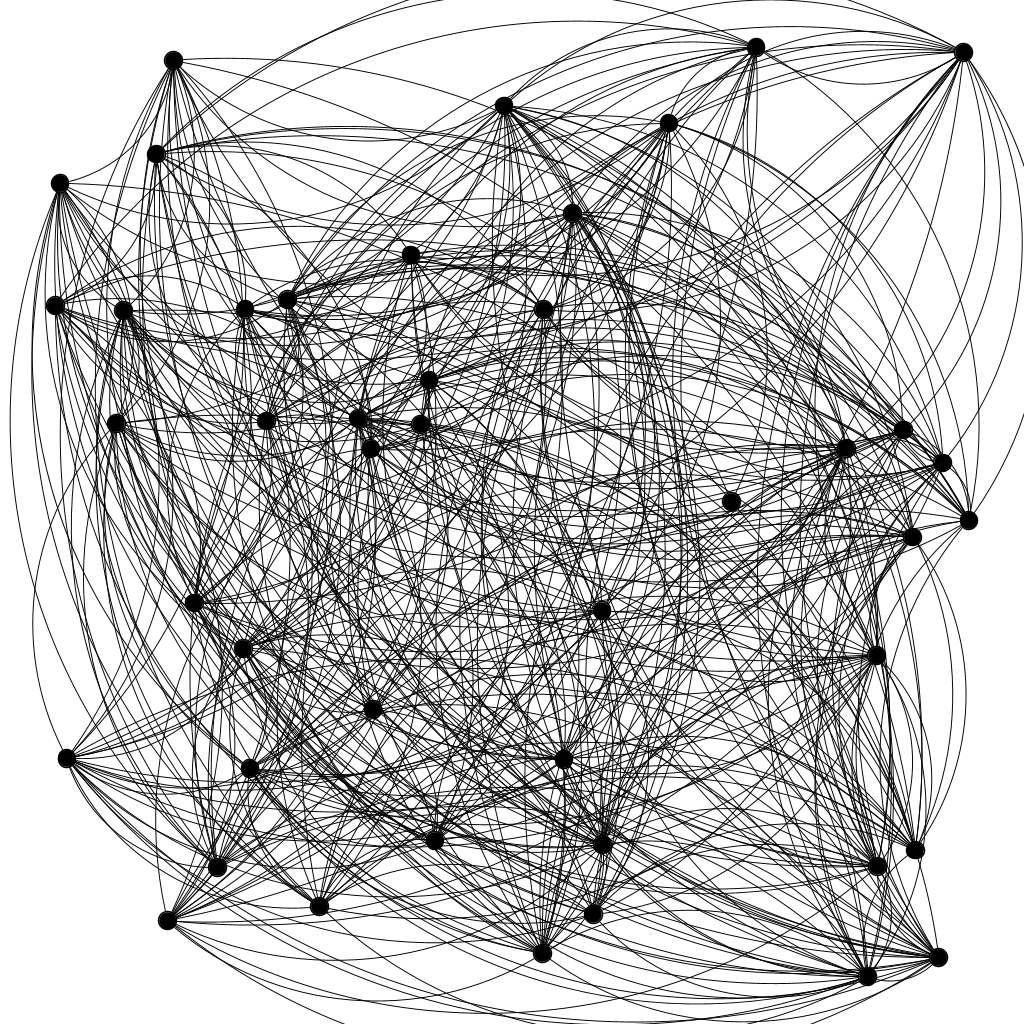}
}
\subfigure(b){
\includegraphics[scale=0.13]{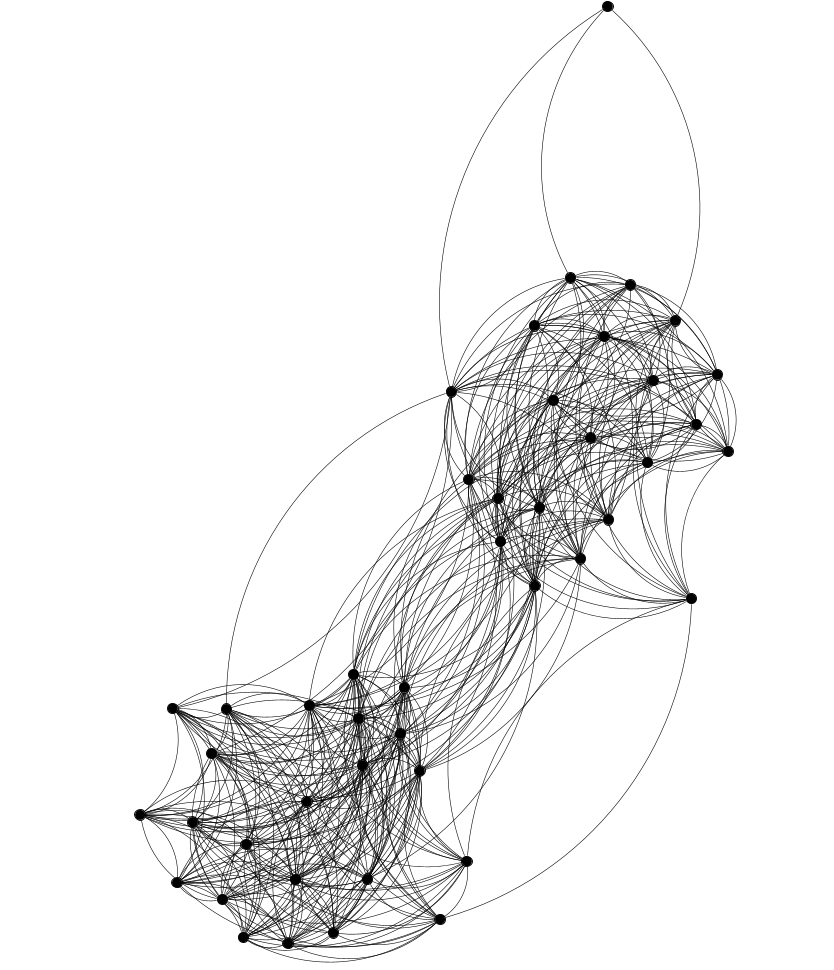}
}

\subfigure(c){
\includegraphics[scale=0.18]{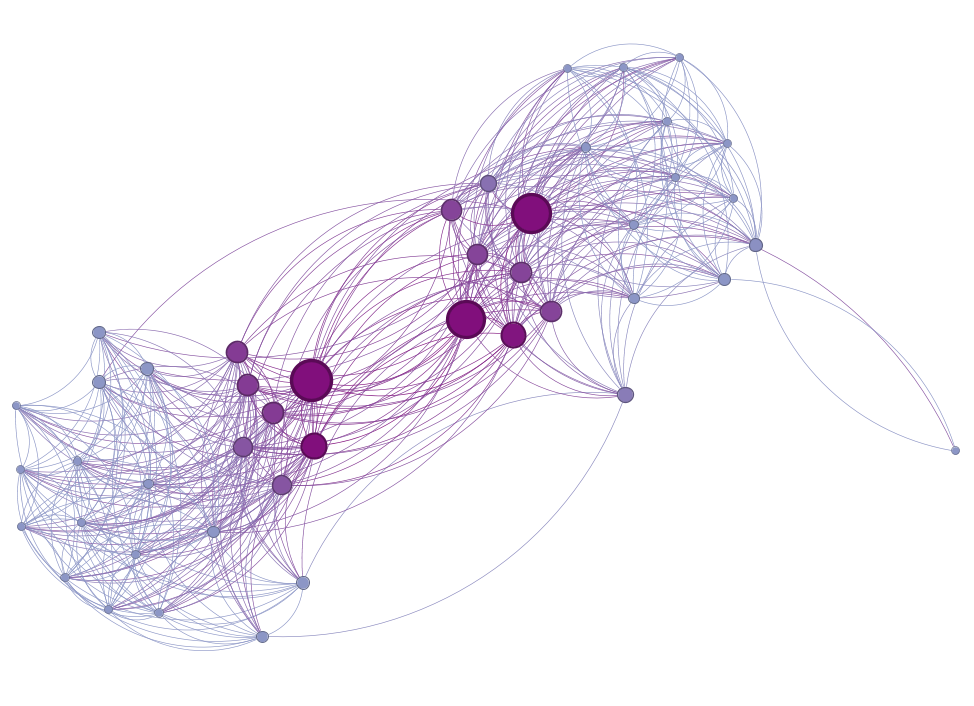}
}
\caption{MCCN Visualizing process for song "550W/Moss". (a) The initial graph of MCCN, which visualizes the music data. (b) The updated graph of MCCN by using the Yifan Hu layout algorithm. (c) The militarized music network, which ranks the nodes and Lines by betweenness centrality. We first update the music network layout by using an algorithm. And then, we calculate betweenness centrality for all nodes and rank the nodes, which is the process from (b) to (c). Nodes that are darker and bigger than others have more relations and similarities with other clips in this music. Also, we can regard the core nodes as the commander of this music because this clip has the most betweenness centrality in the whole music. We can observe that the network for this piece of music exhibits a distinct hierarchical structure. This hierarchy is visually evident through variations in node depth, size, and arrangement. Furthermore, within this musical composition, three segments can be considered relatively significant, as they correspond to the three largest nodes in the network. Additionally, these three segments in the composition exhibit a higher degree of connectivity with other segments. For example, they share similarities in terms of pitch or melody with other segments.} 
\end{figure}

\subsection{Military System Network Case Verification}

After constructing the music network, we employ graph neural network techniques to compare the military strategy network and the music structure network. Before delving into this comparison, let's first gain an understanding of the four military strategy networks: RTN, RAN, SOS, and BA.

The RTN \cite{mn1} creates a linked undirected tree network by randomly selecting existing nodes to establish edges. This network exhibits multiple branches extending from the central node, making it suitable for military operational requirements where resources are distributed evenly, and information flow does not necessitate extensive transit. RTN shares similarities with the \textbf{branch strategy}, executing multiple tasks or activities simultaneously. Additionally, it is akin to the battle tactic of "dividing the army into six separate routes and marching separately" and the metaphor of the "cunning rabbit with three caves."

The RAN \cite{mn1} consists of n nodes generated randomly, initially forming a triangle and then gradually adding nodes. When a new node is introduced, it selects an existing edge randomly, places a new node at its center, and connects the two ends of that edge with the new node. A RAN network features more sub-central points, in addition to the central point, compared to an RTN network. RAN is associated with the \textbf{independent decision-making strategy}, which has historical connections to political institutions such as the feudal system and the military strategy mentioned in Sun Tzu's Art of War, where "soldiers respond randomly to external battles, and in some cases, certain orders from the leader may not be accepted."

\begin{figure}
\centering
\subfigure(a){
\includegraphics[scale=0.17]{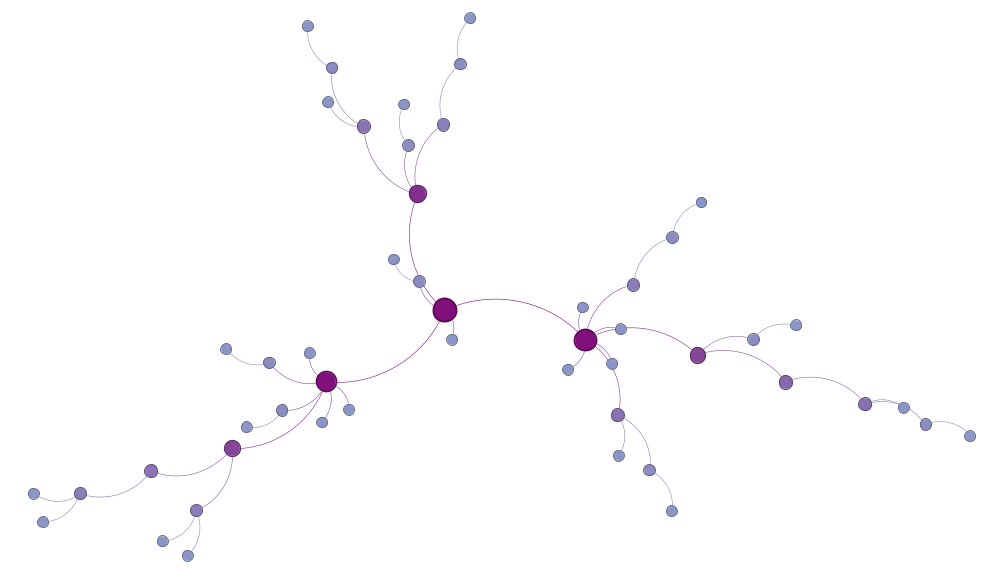}
}

\subfigure(b){
\includegraphics[scale=0.12]{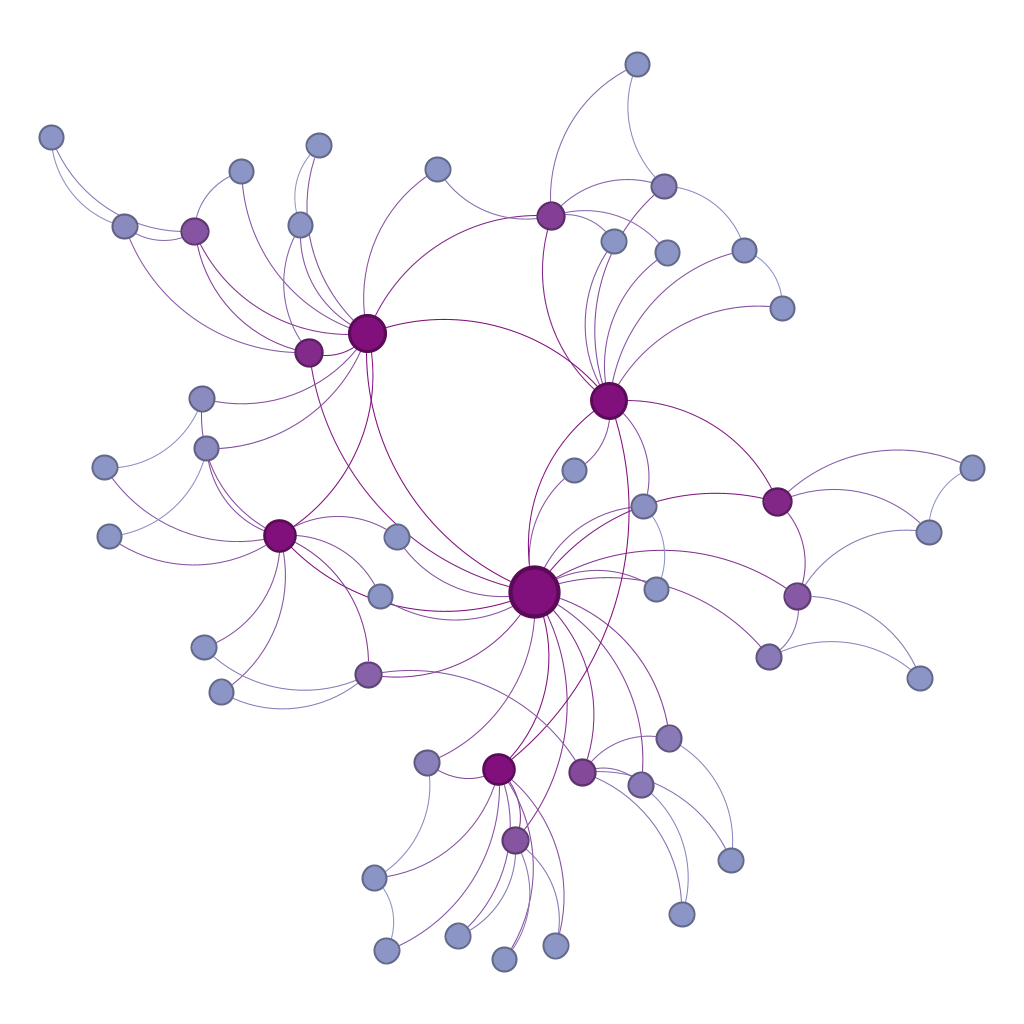}
}

\caption{(a) Random Tree Network and (b) Random Apollo Network which both include 50 nodes.}

\label{figure}
\end{figure}

SOS \cite{mn2} is a multi-level hierarchical randomly generated network characterized by a multi-layer structure, where each node is interconnected by numerous paths, resulting in a higher degree of fault tolerance. SOS is linked to the \textbf{distributed command strategy}, facilitating collaborative subsystems through an efficient command and control framework. The SOS network can be employed for combined operations or the strategy of "making simultaneous frontal and rear attacks," as mentioned in The Thirty-Six Stratagems.

In BA-NW-C2NM \cite{mn2} , each new node is assigned m edges, with each edge having a fixed probability of connecting to an existing node or a random node. This network is well-suited for military operations that necessitate rapid information flow and autonomous collaboration. It is akin to a \textbf{real-time decision-making strategy}, as it receives, disseminates, and processes information in real-time to enable immediate actions. The BA network's information transmission and decision-making occur in a timely manner, resembling collective penetration and defense in military operations.

\begin{figure}
\centering
\subfigure(a){
\includegraphics[scale=0.13]{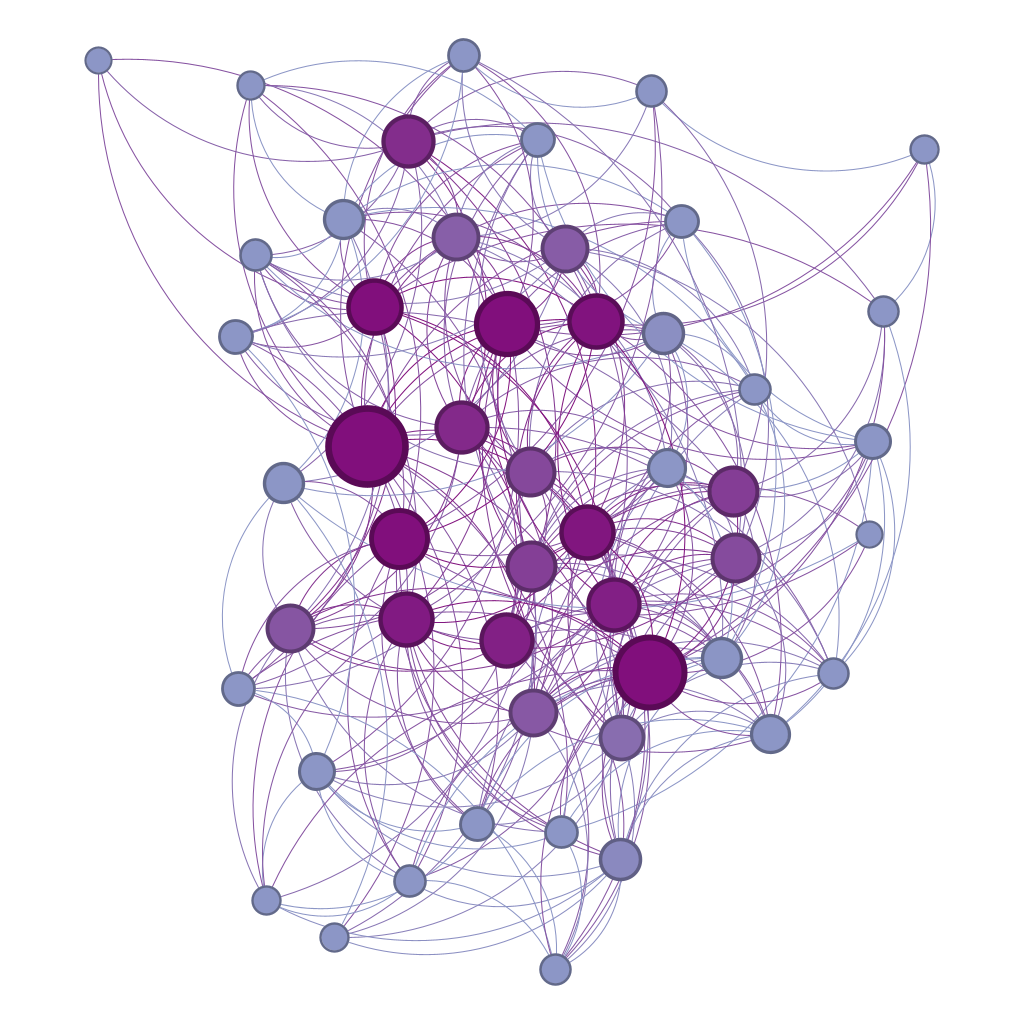}
}

\subfigure(b){
\includegraphics[scale=0.16]{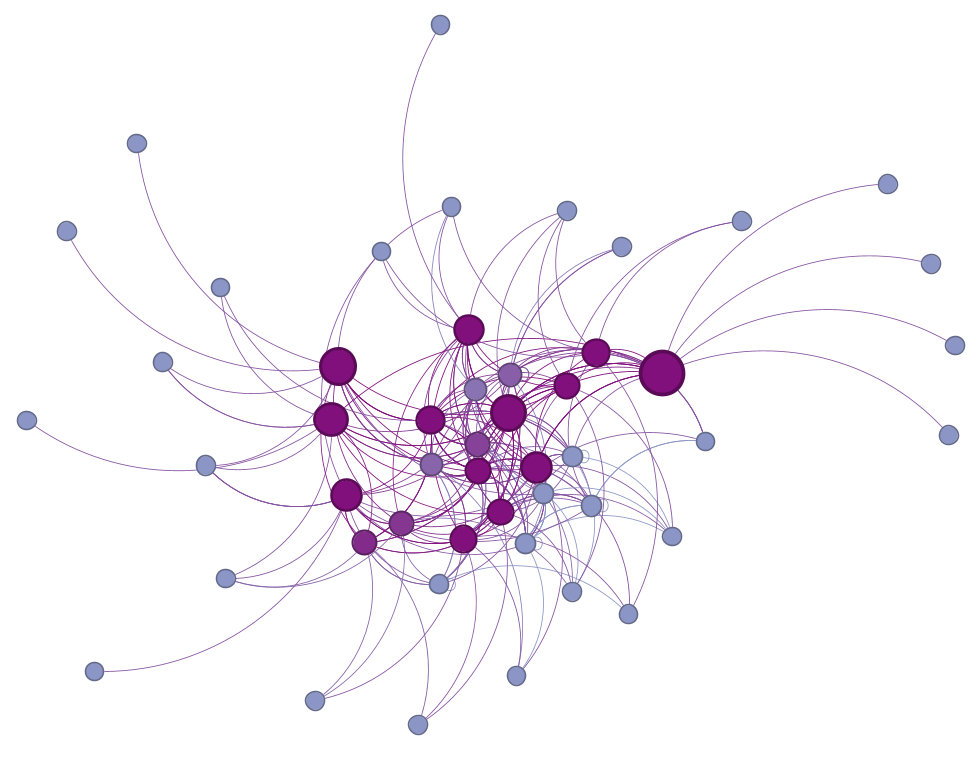}
}

\caption{Graph (a) is the System of System Network and BA-NW-C2NM Network is the (b) graph, which both include 50 nodes.}

\label{figure}
\end{figure}

After constructing the four military strategy networks using their respective algorithms, we employ various standard graph neural network metrics to measure their similarities. This helps us to gain insights into the music systematic features presented by the MCCN network. We assessed four network characteristics: information transmission efficiency, node association, node combination, and network completeness, and computed five characteristic values for each network. A lower dissimilarity value indicates a higher degree of similarity between the two networks.

\textbf{APL (Average Path Length)}: The average length of the shortest path between any two network nodes. It determines the average number of steps required to transfer data across a network or from one node to another. The shorter the average path length, the closer the connection between the nodes and the faster the information dissemination.

\begin{equation}
    APL = \frac{1}{n(n-1)}{\textstyle \sum_{i\ne j}^{d_{ij}}} 
\end{equation}

\textbf{Network Diameter (ND)}: The shortest route length between the network's two farthest nodes. The network diameter is the maximum distance over which information may be disseminated in the network. 
\begin{equation}
    ND = \max{d_{i,j}} 
\end{equation}

\textbf{Graph Density (GD)}: The graph density of a network reflects how tightly nodes are linked to one another. A high density indicates that the nodes are more connected to each other, while a low density indicates that the nodes are more sparsely connected to each other. (m is the number of potential network edges)
\begin{equation}
    GD = \frac{2m}{n(n-1)} 
\end{equation}

\textbf{Modularity (M)}: It's used to calculate the network's community structure. (s is half the network's total number of edges; $k_i$ and $k_j$ represent node i and j's degrees; $\delta (C_i,C_j)$ is an indicator function that is 1 when node i and node j belong to the same community and 0 otherwise)
\begin{equation}
    M = \frac{1}{2S}  {\textstyle \sum_{i,j}^{(A_{i,j} - \frac{k_{i} k_{j}  }{2S} )}} \delta (C_{i}, C_{j})
\end{equation}

\textbf{Clustering Coefficient (CC)}: The clustering coefficient is the ratio of the number of edges that exist between a node's neighbors to the number of feasible edges. A higher clustering coefficient means that there are more tight connections between nodes in the network. ($t_i$ is the amount of connections between a node's neighbor nodes; $k_i$ is the node's degree)
\begin{equation}
    CC = \frac{2t_{i} }{k_{i}(k_{i} - 1) } 
\end{equation}

The dissimilarity score is then computed as follows: ($\omega_{k}$ weight coefficient; $F_{k}^{mus}$ music network (MCCN) feature; $F_{k}^{mil}$ military network feature)
\begin{equation}
    D_{weighted}^{mus, mil} = \sum_{k=1}^{5} \omega_{k} \cdot \left | F_{k}^{mus} - F_{k}^{mil}  \right | 
\end{equation}

After computing these variables, we can calculate the dissimilarity values of the two networks. The selection of weight factors takes into account both the network's military properties and data standardization.

\section{Experiment and Result}

We selected research data from war film soundtracks with high Douban ratings. Out of approximately 300 war films, we had about 2000 music clips to build MCCN. These soundtracks were categorized into two groups: offensive and defensive, each containing 30 songs, based on different movie scenarios. The music used in this research ranges in duration from 2 to 6 minutes. This selection method was employed to avoid the signaling impact of excessively short music pieces, as well as the potential loss of depth in longer songs.

To provide a more comprehensive description of the entire musical piece, we divided it into clips every 6 seconds. This approach ensures that the number of network nodes for the shortest duration of music remains adequate and that the richness of each node is generally consistent. Furthermore, we took measures to exclude any music with vocal or lyrical components to eliminate the influence of voice and lyrics on the study.

When selecting \textbf{defensive music}, we considered film sequences depicting war failure, sadness, escape, rescue, and sacrifice. In contrast, \textbf{offensive music} was chosen based on war attack, victory, occupation, and the battle for victory. Each category of music exclusively reflects the emotion associated with that category, with no mixture of emotions allowed. For instance, offensive music cannot contain elements of both one side attacking and the other side retreating.

\begin{figure}
\centering
\subfigure(a){
\includegraphics[scale=0.12]{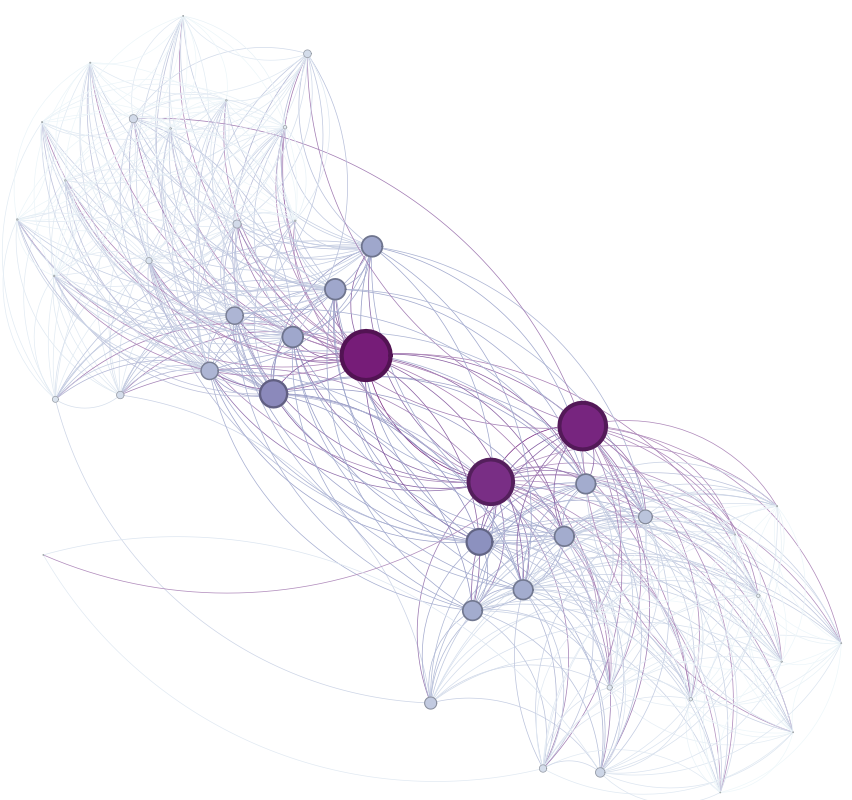}
}
\subfigure(b){
\includegraphics[scale=0.14]{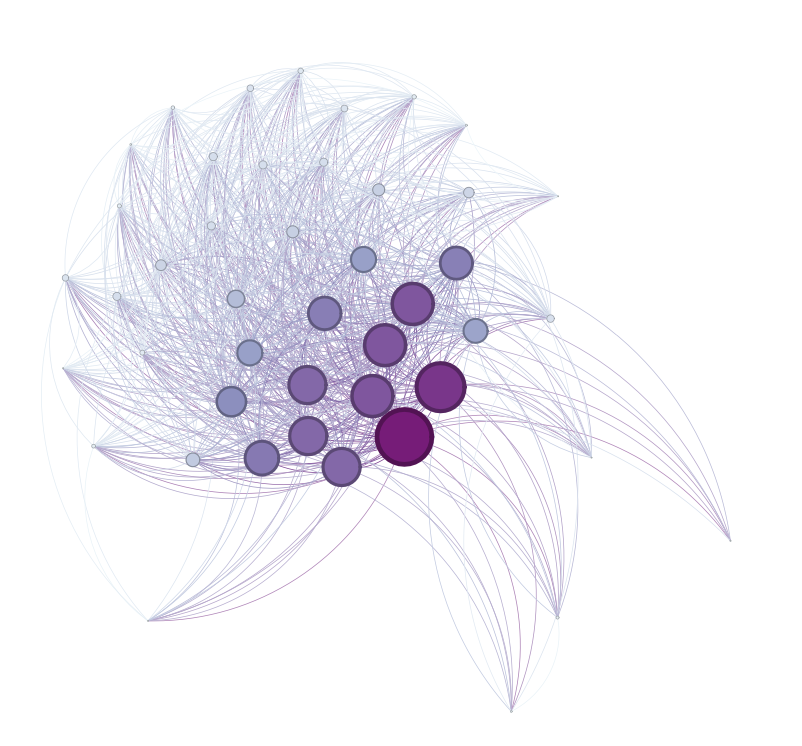}
}

\subfigure(c){
\includegraphics[scale=0.13]{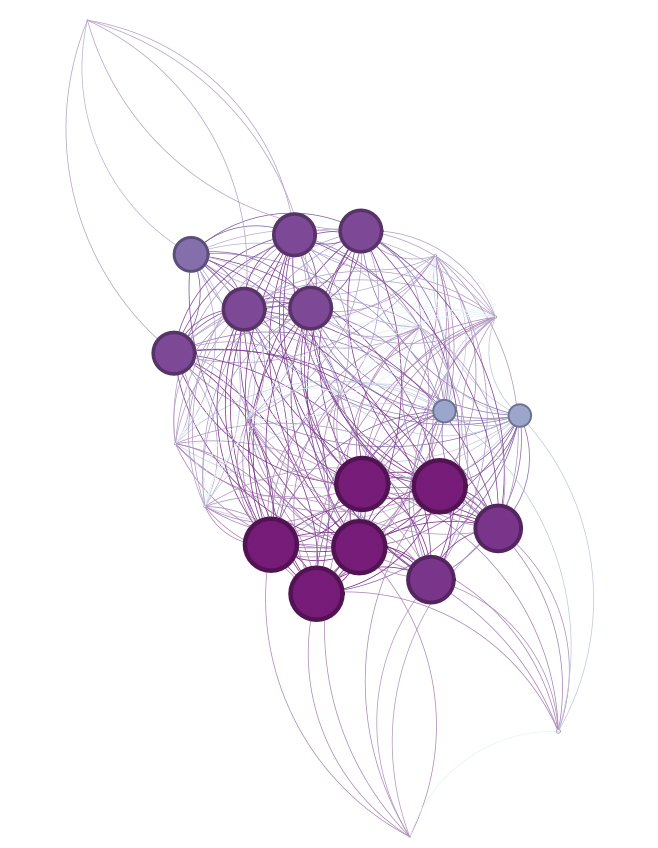}
}
\subfigure(d){
\includegraphics[scale=0.16]{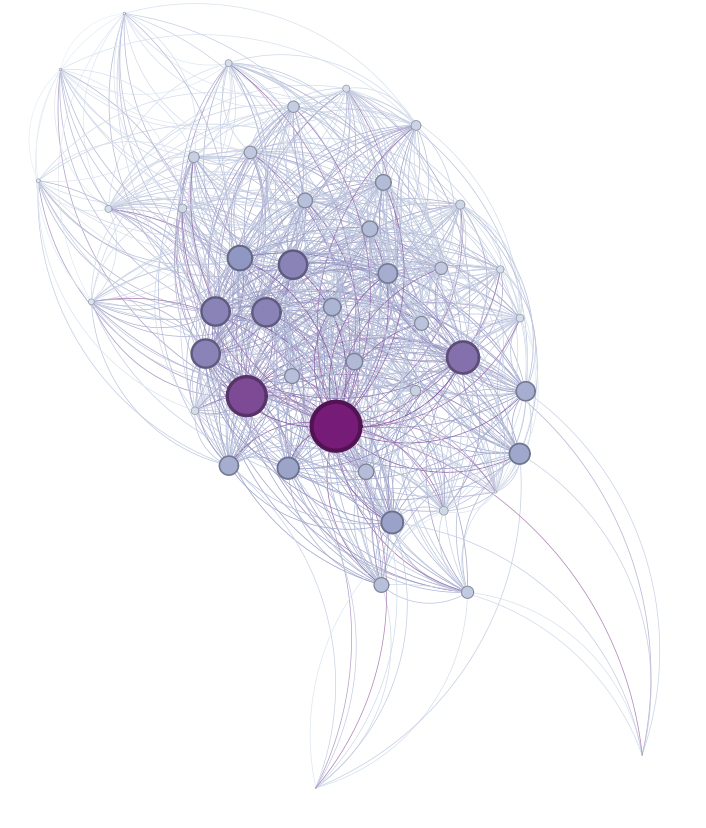}
}
\caption{(a) and (b) are offensive MCCNs, while (c) and (d) are defensive MCCNs. Offensive MCCNs have obvious hierarchy, while nodes in defensive MCCNs almost have the same betweenness centrality.}
\label{f4}
\end{figure}

After grouping and calculating the music based on the corresponding film scenes and constructing the MCCN for 60 songs, we proceeded to observe the MCCN for offensive and defensive music categories. Analyzing these two categories of MCCN based on node distribution, network hierarchy, and visual patterns, we identified four typical MCCNs, as shown in figure \ref{f4}. We assigned weight coefficients to the two categories of music based on the elements of offense and defense in military actions, as presented in table \ref{t1}. Using these weight coefficients, we calculated the similarity between the MCCN and the military action network, as detailed in table \ref{t2}.

Based on our findings and result in table \ref{t3}, there are three significant similarities between the MCCNs and the military networks:

a) The offensive MCCN exhibits lower similarity compared to the defensive MCCN. This suggests that offensive music is inherently more intricate and challenging to analogize to a single military network.

b) Among the military networks, the offensive MCCN bears the highest resemblance to the SOS network, while the defensive MCCN aligns most closely with the BA network.

c) The disparities among the four similarity values highlight that the offensive MCCN shares a higher similarity score with both the SOS and BA networks, with a more pronounced gap from the other two military networks. This implies that, aside from the SOS network, the offensive MCCN also possesses certain characteristics akin to the BA network. On the other hand, the defensive MCCN exhibits dissimilarity scores that are somewhat comparable to the BA, SOS, and RAN military networks, indicating that it incorporates some features from both the SOS and RAN networks.

\begin{table}
\centering
\caption{The weighted coefficients for offense and defense groups.}\label{t1}
\begin{tabular}{|l|l|l|l|l|l|}
\hline
 & APL & ND & GD & M & CC \\
\hline
Offense & 0.4 & 0.4 & 0.05 & 0.075 & 0.075 \\
\hline
Defense & 0.05 & 0.075 & 0.4 & 0.075 & 0.4 \\
\hline
\end{tabular}
\end{table}

\begin{table}
\centering
\caption{The dissimilarity value between military network and MCCNs. The bigger the value is, the less similarity the two network will have.}\label{t2}
\begin{tabular}{|l|l|l|l|l|}
\hline
 &  RTN & RAN & SOS & BANWC2NM \\
\hline
Offense1 & 2.8192 & 0.9701 & 0.2879 & 0.3732 \\
\hline
Offense2 & 4.7982 & 1.8131 & 0.7933 & 1.0656 \\
\hline
Defense1 & 1.1497 & 0.4743 & 0.4187 & 0.3179 \\
\hline
Defense2 & 1.1148 & 0.4570 & 0.3972 & 0.3111 \\
\hline
\end{tabular}
\end{table}

\begin{table}
\centering
\caption{The dissimilarity value for offensive and defensive MCCNs. The bigger the value is, the less similarity the two network will have.}\label{t3}
\begin{tabular}{|l|l|l|l|l|}
\hline
 &  RTN & RAN & SOS & BANWC2NM \\
\hline
Offense & 3.8608 & 1.4138 & 0.5539 & 0.7377 \\
\hline
Defense & 1.1368 & 0.4679 & 0.4108 & 0.3154 \\
\hline
\end{tabular}
\end{table}

\section{Discussion}

From the viewpoint of music performance, offensive music serves an inspiring role, characterized by its emotionally charged and pronounced dynamics. On the other hand, defensive music typically maintains a steady rhythm to complement the corresponding defensive or withdrawal scenes in film, often with less pronounced melodic variations. The experimental results validate these observations. By systematically showcasing music structure, we find that offensive MCCN is most similar to the SOS network, both exhibiting clear hierarchical structures. In contrast, defensive MCCN shares similarities with the BA, SOS, and RAN networks, with high similarity between individual music segments and no particularly pronounced variations.

Analyzing from the perspective of military exercises, offensive military operations typically involve different roles and objectives for various parts of the army. Roles such as commanders, sentinels, vanguards, rearguards, and generals have distinct hierarchical planning. Similarly, offensive music structures also exhibit clear hierarchical arrangements. In contrast, during defensive and withdrawal operations, different parts of the army often share the same objectives, such as preserving military strength and ensuring a safe retreat for everyone.

\begin{figure}
    \centering
    \includegraphics[scale=0.3]{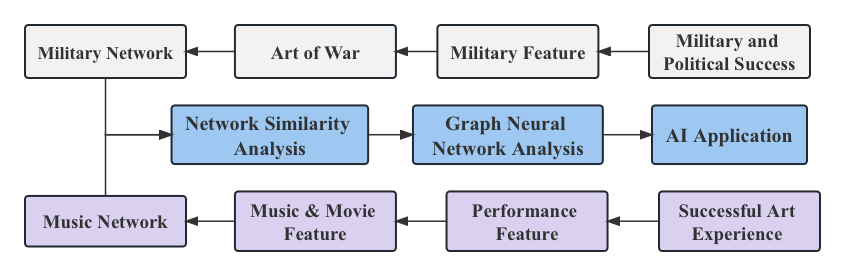}
    \caption{The figure shows the relationship between music structure and military strategy. When exploring the similarities between music structure and military strategies, we evaluate the resemblance of neural networks built from data modeling. The objectives of "making music more pleasing" and "devising a successful military strategy" share a common theme: the pursuit of enabling a system to possess a relatively perfect network structure. Crucial to this endeavor is the careful selection of the size of each data node, ensuring a balanced relationship among nodes within the system. By analogy to real-world society, when each subsystem maintains a well-balanced relationship within the overall system, we are better equipped to establish a more robust societal system.}
    \label{f5}
\end{figure}

In summary, from the perspectives of music perception and military exercises, the "systemic nature of music" presented by MCCN can be interpreted. As shown in Figure \ref{f5}, we have used AI methods to connect two art forms, both possessing "systemic nature," and visually present music with different characteristics using a systematic visualization approach. By generating and observing the MCCN of music, we can visually perceive the overall characteristics of a piece of music. Since each node in the MCCN contains all the music features of its corresponding segment (pitch, loudness, etc.), we can see the connections between different segments of a piece of music within the MCCN. Moreover, we can use this model to quickly extract the essence clip of music (segments that have higher cosine similarity with other segments), thereby gaining a better understanding and perception of music.

\section{Conclusion}

This research aims to propose a novel approach for the systematic perception of music through the establishment of the MCCN network model. Furthermore, it connects the principles of Sun Tzu's Art of War and political institutions with military system network models. The alignment of political institutions and strategies with data models enhances our understanding of how social systems operate. Drawing parallels between the musical scores of conductors and the sandbox exercises of military commanders, these analyses reveal potential connections regarding "systemic" aspects between music structure and military strategies. This provides a fresh perspective for music perception and aesthetic education. In the future, we aspire to delve into more comprehensive music perception through the coordination of multiple senses and develop more precise systemic perception models.

% \begin{table*}
% \centering
% \caption{A specific assessment of the network coefficients of the 50 offensive and defensive soundtracks.}\label{t1}
% \begin{tabular}{llllll}
% \toprule
%  & Average Path Length& Network Diameter& Graph Density& Modularity& Clustering Coefficient\\
% \midrule
% Offense & 0.4 & 0.4 & 0.05 & 0.075 & 0.075 \\

% Defense & 0.05 & 0.075 & 0.4 & 0.075 & 0.4 \\
% \bottomrule 
% \end{tabular}
% \end{table*}

\bibliographystyle{IEEEtran}
\bibliography{reference.bib}
\nocite{*}

\end{document}